\documentstyle[aps]{revtex}

\begin{document}
\title{Parametrizing the Neutrino Mixing Matrix}
\author{A. Zee}
\address{Kavli Institute for Theoretical Physics\\
University of California\\
Santa Barbara, CA 93106\\
USA\\
zee@kitp.ucsb.edu}
\maketitle

\begin{abstract}
We propose parametrizations of the neutrino mixing matrix. We discuss
several Ansatzes: in some of them $V_{e3}=0$ but in others $V_{e3}$
typically ``comes out'' to be of order $\simeq 0.05.$
\end{abstract}

\bigskip

{\bf {\large I. Neutrino Mixing Matrix}}

Wolfenstein's parametrization\cite{wolfpara} of the quark mixing matrix,
while not based on any deep theoretical foundation, has proved to be
enormously useful in making sense of the data. Something similar is needed
to make sense of the neutrino mixing matrix. In reporting the value of a
typical mixing angle $\theta ,\ $experimentalists quote variously $\sin
\theta ,$ $\sin 2\theta ,$ $\tan \theta ,$ or their squares. It would be
helpful to have a more systematic way of quoting experimental values,
analogous to what the Wolfenstein parametrization has provided for the quark
sector. In the literature, two parametrizations\cite{bkaus,xingpara} of
neutrino mixing had been proposed that I know of. Here we would to propose a
quite different parametrization. We also suggest a more systematic way for
experimentalists to quote their data. As in the Wolfenstein parametrization,
our discussion is not based on any deep theoretical foundation, but rather
on a phenomenological analysis of the available data. This paper is thus a
somewhat old fashioned mix of Ansatz making and data fitting.

\medskip Thanks to heroic experimental efforts, the neutrino mixing matrix
is now known to take on the value 
\begin{equation}
|V|=\left( 
\begin{array}{lll}
0.72-0.88 & 0.46-0.68 & <0.22 \\ 
0.25-0.65 & 0.27-0.73 & 0.55-0.84 \\ 
0.10-0.57 & 0.41-0.80 & 0.52-0.83
\end{array}
\right)  \label{exp}
\end{equation}
I am citing the report\cite{GonGar} of Gonzalez-Garcia, based in part on the
analysis given by Bachall, Gonzalez-Garcia, and Pena-Garay\cite{bgp}. Here
the notation $|V|$ is such that the $ij$ matrix element of the matrix $|V|$
is equal to $|V_{ij}|.$

The mixing matrix $V$ relates the neutrino current eigenstates (denoted by $%
\nu _{\alpha }$ ($\alpha =e,$ $\mu ,$ $\tau )$ and coupled by the $W$ bosons
to the corresponding charged leptons) to the neutrino mass eigenstates
(denoted by $\nu _{i}$ ($i=1,2,3))$ according to 
\begin{equation}
\left( 
\begin{array}{l}
\nu _{e} \\ 
\nu _{\mu } \\ 
\nu _{\tau }
\end{array}
\right) =V\left( 
\begin{array}{l}
\nu _{1} \\ 
\nu _{2} \\ 
\nu _{3}
\end{array}
\right)
\end{equation}

We will take the neutrinos to be Majorana\cite{kay} as seems likely, so that
we have in the Lagrangian the mass term 
\begin{equation}
{\cal L}=-\nu _{\alpha }M_{\alpha \beta }C\nu _{\beta }+h.c.
\end{equation}
where $C$ denotes the charge conjugation matrix. Thus, the neutrino mass
matrix $M$ is symmetric.

For the sake of simplicity we will assume $CP$ conservation so that $M$ is
real. With this simplification, $M$ is diagonalized by an orthogonal
transformation 
\begin{equation}
V^{T}MV=\left( 
\begin{array}{lll}
m_{1} & 0 & 0 \\ 
0 & m_{2} & 0 \\ 
0 & 0 & m_{3}
\end{array}
\right)  \label{diag}
\end{equation}
It will become clear to the reader that everything in this paper is easily
generalized to the case in which $CP$ is violated so that $V$ is unitary
rather than orthogonal, as briefly indicated in section V. Also, we are free
to multiply $V$ on the right by some diagonal matrix whose diagonal entries
are equal to $\pm 1$. This merely multiplies each of the columns in $V$ by
an arbitrary sign. Various possible phases have been discussed in detail in
the literature\cite{nieves,sign}.

\smallskip

{\bf {\large II. An Ansatz for the Neutrino Mixing Matrix}}

\smallskip

We could suppose either that the entries in $V$ represent a bunch of
meaningless numbers (for example, possibly varying from domain to domain in
the universe) or that they point to some deeper structure or symmetry. In
the latter spirit, let us make a guess of what $V$ might be.

Since $V_{e3}$ appears to be small, let us boldly set it to $0$. Next, since 
$1/\sqrt{2}\sim 0.707$ we will guess that $V_{\mu 3}=(0.55-0.84)=1/\sqrt{2}.$
Finally, since $1/\sqrt{3}\sim 0.577$ we will set $V_{e2}=(0.46-0.68)=1/%
\sqrt{3}.$ In other words, we propose that we know the upper triangular
entries of the matrix $V:$

\begin{equation}
V=\left( 
\begin{array}{rrr}
{X} & {\frac{1}{\sqrt{3}}} & 0 \\ 
{X} & {X} & {\frac{1}{\sqrt{2}}} \\ 
{X} & {X} & {X}
\end{array}
\right) ,  \label{fir}
\end{equation}
where ${X}$ denotes an unknown quantity.

Remarkably, this essentially fixes the mixing matrix $V$. Once we take the
last column to be proportional to $(0,1,-1)$, orthogonality and our
``knowledge'' that $V_{e2}$ is $1/\sqrt{3}$ immediately fix the second
column to be proportional $(1,1,1)$ and hence the first column to be
proportional to $(-2,1,1)$. X. G. He and I therefore arrived at the Ansatz
or guess\cite{hz1} 
\begin{equation}
V=\left( 
\begin{array}{rrr}
-{\frac{2}{\sqrt{6}}} & {\frac{1}{\sqrt{3}}} & 0 \\ 
{\frac{1}{\sqrt{6}}} & {\frac{1}{\sqrt{3}}} & {\frac{1}{\sqrt{2}}} \\ 
{\frac{1}{\sqrt{6}}} & {\frac{1}{\sqrt{3}}} & {-\frac{1}{\sqrt{2}}}
\end{array}
\right) .  \label{theV}
\end{equation}
This Ansatz was proposed earlier by Harrison, Perkins and Scott\cite{hps}.
Also, this mixing matrix (but curiously, with the first and second column
interchanged) was first suggested by Wolfenstein more than 20 years ago\cite
{wolf78} based on some considerations involving the permutation group $S_{3}$%
. It has subsequently been studied extensively by Harrison, Perkins and Scott%
\cite{21}, and by Xing\cite{22}. Attempts to derive this mixing matrix have
been discussed by Low and Volkas\cite{lv}. We will refer to the matrix in (%
\ref{theV}) as $V$ unless confusion could arise in which case we will refer
to it as $V_{HPSHZ.}.$

The signs in $V$ were chosen to minimize the number of minus signs. The
vanishing of $V_{e3}$ indicates, if this Ansatz is correct, that there is no
observable $CP$ violating effects in neutrino oscillation experiments. See (%
\ref{cklike}) below.

The mixing matrix $V$ may be factorized as 
\begin{equation}
V=V_{23}V_{12}  \label{fac}
\end{equation}
where 
\begin{equation}
V_{23}=\left( 
\begin{array}{rrr}
1 & {0} & 0 \\ 
{0} & {\frac{1}{\sqrt{2}}} & {\frac{1}{\sqrt{2}}} \\ 
{0} & {\frac{1}{\sqrt{2}}} & {-\frac{1}{\sqrt{2}}}
\end{array}
\right)
\end{equation}
and 
\begin{equation}
V_{12}=\left( 
\begin{array}{rrr}
-\sqrt{\frac{2}{3}} & {\frac{1}{\sqrt{3}}} & 0 \\ 
{\frac{1}{\sqrt{3}}} & \sqrt{\frac{2}{3}} & {0} \\ 
{0} & {0} & {1}
\end{array}
\right) .
\end{equation}
Note that our sign choices in (\ref{theV}) is such that $\det V_{23}=\det
V_{12}=-1$ (but $\det V=1.)$

The rotation of the neutrino current eigenstates appear to ``occur in two
steps.'' In other words, if we follow Wolfenstein and define $\nu _{x}\equiv
(\nu _{\mu }+\nu _{\tau })/\sqrt{2}$ and $\nu _{y}\equiv (\nu _{\mu }-\nu
_{\tau })/\sqrt{2}$ we find that the mass eigenstates are given by 
\begin{equation}
\nu _{1}=-\sqrt{\frac{2}{3}}\nu _{e}+{\frac{1}{\sqrt{3}}}\nu _{x},
\end{equation}
\begin{equation}
\nu _{2}={\frac{1}{\sqrt{3}}}\nu _{e}+\sqrt{\frac{2}{3}}\nu _{x},
\end{equation}
and 
\begin{equation}
\nu _{3}=\nu _{y}
\end{equation}
The matrix $V_{12}$ describes a rotation through $\arcsin ({\frac{1}{\sqrt{3}%
})}\sim 35^{o}.$

Giunti\cite{giunti} has proposed another Ansatz, to which we will turn later.

\bigskip

{\bf {\large III. Parametrizing the Neutrino Mixing Matrix }}

\smallskip

In order to compare theory with experiment, we have to understand what the
experimental values in (\ref{exp}) mean. The quoted range of numbers are
extracted from different experiments, but the error ranges are obviously not
uncorrelated since $V$ has to be unitary, or orthogonal with our simplifying
assumptions.

There is not a unique way to extract an orthogonal matrix $V_{\exp }$ from (%
\ref{exp}). Of the many possible ways we choose the following procedure. We
take the mean value of the cited range of the various entries in $|V|$ (e.g.
for $|V_{e2}|$ we set $(0.46-0.68)$ $\rightarrow 0.57$ ), fix the signs of
various entries to agree with the choices in (\ref{theV}), and thus arrive
at the ``experimental mean values'' 
\begin{equation}
V_{\text{mean}}=\left( 
\begin{array}{lll}
-0.8 & 0.57 & 0.00 \\ 
0.45 & 0.5 & 0.695 \\ 
0.335 & 0.605 & -0.675
\end{array}
\right)  \label{mean}
\end{equation}

\smallskip The matrix $V_{\text{mean}}$ is not orthogonal. Since $V_{\text{
mean}}^{T}V_{\text{mean}}$ is real symmetric it can be diagonalized by an
orthogonal matrix $S$ so that the matrix $K^{2}\equiv S^{T}V_{\text{mean}
}^{T}V_{\text{mean}}S$ is diagonal with positive definite diagonal elements.
We then define the ``corrected experimental mixing matrix'' to be the
orthogonal matrix 
\begin{equation}
V_{\text{exp}}^{\text{corr}}\equiv V_{\text{mean}}(SK^{-1}S^{T}).
\label{correxp}
\end{equation}
Numerically, this comes out to be 
\begin{equation}
V_{\text{exp}}^{\text{corr}}=\left( 
\begin{array}{lll}
-0.814 & 0.578 & 0.056 \\ 
0.436 & 0.545 & 0.716 \\ 
0.384 & 0.607 & -0.696
\end{array}
\right)  \label{correxpnum}
\end{equation}
Note that a non-zero value of $V_{e3}$ is ``generated'' but one order of
magnitude smaller than the other entries. Second, note that $V_{\text{exp}}^{%
\text{corr}}$ is well within the range quoted in (\ref{exp}). Finally,
compare $V_{\text{exp}}^{\text{corr}}$ with the theoretical Ansatz (\ref
{theV}) 
\begin{equation}
V=\left( 
\begin{array}{lll}
-0.817 & 0.577 & 0.000 \\ 
0.408 & 0.577 & 0.707 \\ 
0.408 & 0.577 & -0.707
\end{array}
\right)  \label{theVnum}
\end{equation}

Clearly, $V$ and $V_{\text{exp}}^{\text{corr}}$ are rather close. Thus,
three obvious ways of parametrizing the data suggest themselves. We could
parametrize the data by a matrix $W$ defined by $V_{\text{exp}}^{\text{corr}%
}=VW,$ or a matrix $X$ defined by $V_{\text{exp}}^{\text{corr}}=XV,$ or a
matrix $Y$ defined by $V_{\text{exp}}^{\text{corr}}=V_{23}YV_{12}$ (as
suggested by (\ref{fac})). Namely, we parametrize the deviation of $V$ from $%
V_{\text{exp}}^{\text{corr}}.$

In each of these three possible ways, the matrix $(W,$ or $X,$ or $Y)$ will
be close to the identity matrix, and we could then follow Wolfenstein and
parametrize them accordingly. In each of these cases, we will write the
orthogonal matrix $W,$ or $X,$ or $Y$ in the form $e^{A}\simeq I+A+\frac{1}{2%
}A^{2}$ with 
\begin{equation}
A=\left( 
\begin{array}{lll}
0 & \alpha & -\gamma \\ 
-\alpha & 0 & \beta \\ 
\gamma & -\beta & 0
\end{array}
\right)  \label{A}
\end{equation}
We will discuss each of these possibilities in turn.

\bigskip (I) We solve the 3 equations $W_{jj}=(V^{T}V_{\text{exp}}^{\text{
corr}})_{jj},$ $j=1,2,3$ (no repeated summation convention) for the 3
unknowns $\alpha ,\beta ,$ and $\gamma .$ To the order indicated, we have 3
coupled quadratic equations and we obtain 8 solutions corresponding to
various choices of signs. We then compute $\sum_{i\neq j}(W_{ij}-(V^{T}V_{%
\text{exp}}^{\text{corr}})_{ij})^{2}$ and determine the solution which
minimizes this quantity. In this way, we find that the best fit to $W$ is
given by $\alpha =-0.00249,\beta =0.0440,$ and $\gamma =0.0374.$ The main
outcome of this simple phenomenological analysis is that $\alpha $ is an
order of magnitude smaller than $\beta $ and $\gamma ,$ which appear to be
comparable. Indeed, we see that $\alpha $ is close to $-3\beta ^{2}/2.$ In
the spirit of Wolfenstein's paper, without any theoretical understanding
whatsoever of the values of $\alpha ,\beta ,$ and $\gamma ,$ we could try,
instead of a 3-parameter fit, a 2-parameter fit by setting 
\begin{equation}
A=\left( 
\begin{array}{lll}
0 & -3\nu \beta ^{2}/2 & -\beta \\ 
3\nu \beta ^{2}/2 & 0 & \beta \\ 
\beta & -\beta & 0
\end{array}
\right)
\end{equation}
where we expect $\nu $ to be of order 1. Solving the 2 equations $%
W_{12}=(V^{T}V_{\text{exp}}^{\text{corr}})_{12}$ and $W_{23}=(V^{T}V_{\text{%
exp}}^{\text{corr}})_{23},$ we find that $\nu =0.908$ and $\beta =0.0439.$

(II) The analysis clearly proceeds in the same way: we solve the 3 equations 
$X_{jj}=(V_{\text{exp}}^{\text{corr}}V^{T})_{jj}$ and obtain the best fit
solution $\alpha =0.0413,$ $\beta =-0.0143,$ and $\gamma =0.0378.$

(III) In this case we solve the 3 equations $Y_{jj}=(V_{23}^{T}V_{\text{exp}%
}^{\text{corr}}V_{12}^{T})_{jj}$ and obtain the best fit solution $\alpha
=0.00249,$ $\beta =0.0143,$ and $\gamma =-0.0559.$ In other words, we have
to rotate $\nu _{e},$ $\nu _{x}\equiv (\nu _{\mu }+\nu _{\tau })/\sqrt{2}$
and $\nu _{y}\equiv (\nu _{\mu }-\nu _{\tau })/\sqrt{2}$ by a tiny amount
before mixing them with $V_{12}^{T}.$

\bigskip

{\bf {\large IV. Another Ansatz for the Neutrino Mixing Matrix}}

\bigskip

Giunti\cite{giunti} has proposed another interesting Ansatz for the mixing
matrix 
\begin{equation}
V_{G}=\left( 
\begin{array}{rrr}
-{\frac{\sqrt{3}}{2}} & {\frac{1}{2}} & 0 \\ 
{\frac{1}{2\sqrt{2}}} & \frac{\sqrt{3}}{2\sqrt{2}} & {\frac{1}{\sqrt{2}}} \\ 
{\frac{1}{2\sqrt{2}}} & \frac{\sqrt{3}}{2\sqrt{2}} & {-\frac{1}{\sqrt{2}}}
\end{array}
\right)  \label{VGiunti}
\end{equation}
which also factorizes nicely $V_{G}=V_{23}V_{G12}$ where 
\begin{equation}
V_{G12}=\left( 
\begin{array}{rrr}
-{\frac{\sqrt{3}}{2}} & {\frac{1}{2}} & 0 \\ 
{\frac{1}{2}} & {\frac{\sqrt{3}}{2}} & {0} \\ 
{0} & 0 & {1}
\end{array}
\right) .
\end{equation}
Thus, $V_{G}$ and $V$ differ by a rotation of about $5^{o}$ in the $(1-2)$
plane: 
\begin{equation}
V_{G}=V\left( 
\begin{array}{rrr}
{\frac{1}{\sqrt{2}}+\frac{1}{2\sqrt{3}}} & {\frac{1}{2}-}\frac{1}{\sqrt{6}}
& 0 \\ 
-{\frac{1}{2}+}\frac{1}{\sqrt{6}} & {\frac{1}{\sqrt{2}}+\frac{1}{2\sqrt{3}}}
& {0} \\ 
{0} & 0 & {1}
\end{array}
\right) .
\end{equation}
Numerically, 
\begin{equation}
V_{G}=\left( 
\begin{array}{rrr}
-{0.866} & {0.5} & 0 \\ 
{0.354} & {0.612} & {0.707} \\ 
{0.354} & 0.612 & {-0.707}
\end{array}
\right) ,
\end{equation}
which is to be compared to $V$ in (\ref{theVnum}). We see that $V_{G}$ makes
the $\mu 1$ and $\tau 1$ entries smaller ($0.354$ rather than $0.408)$ and
the $\mu 2$ and $\tau 2$ entries larger ($0.612$ rather than 0.577).

\bigskip We could ask whether $V$ or $V_{G}$ is closer to the experimental
mixing matrix $V_{\text{exp}}^{\text{corr}}$. The group invariant distance
between two group elements $O_{1}$ and $O_{2}$ of $SO(3)$ is given by 
\begin{equation}
d_{12}\equiv 3-Tr(O_{1}^{T}O_{2})  \label{dist}
\end{equation}
We find 
\begin{equation}
d(V)\equiv 3-Tr(V^{T}V_{\text{exp}}^{\text{corr}})\simeq 0.00334
\end{equation}
and 
\begin{equation}
d(V_{G})\equiv 3-Tr(V_{G}^{T}V_{\text{exp}}^{\text{corr}})\simeq 0.0122,
\end{equation}
about a factor of $4$ larger than $d(V).$ If (and only if) we believe in the
more or less arbitrary way in which $V_{\text{exp}}^{\text{corr}}$ is
extracted from the raw data (\ref{exp}), then $V$ is a somewhat better fit
to the data than $V_{G}.$

\bigskip

\bigskip

{\bf {\large V. Angular Parametrization of the Neutrino Mixing Matrix}}

\bigskip

\bigskip

\smallskip We could of course always parametrize the neutrino mixing matrix
with three Euler angles and a phase. The factorization in (\ref{fac})
suggests that we use the analog of the Chau-Keung\cite{chaukeung}
parametrization commonly used for the Cabibbo-Kobayashi-Maskawa matrix in
the quark sector and thus we write (where as usual $c_{23}\equiv \cos \theta
_{23},$ $s_{23}\equiv \sin \theta _{23},$ and so forth) 
\begin{eqnarray}
V_{{\rm angular}} &=&V_{23}V_{31}V_{12}  \label{cklike} \\
&=&\left( 
\begin{array}{lll}
1 & 0 & 0 \\ 
0 & c_{23} & s_{23} \\ 
0 & s_{23} & -c_{23}
\end{array}
\right) \left( 
\begin{array}{lll}
c_{31} & 0 & s_{31}e^{-i\phi } \\ 
0 & 1 & 0 \\ 
-s_{31}e^{i\phi } & 0 & c_{31}
\end{array}
\right) \left( 
\begin{array}{lll}
-c_{12} & s_{12} & 1 \\ 
s_{12} & c_{12} & 0 \\ 
0 & 0 & 1
\end{array}
\right) \\
&=&\left( 
\begin{array}{lll}
-c_{31}c_{12} & c_{31}s_{12} & s_{31}e^{-i\phi } \\ 
s_{12}c_{23}+c_{12}s_{23}s_{31}e^{i\phi } & 
c_{12}c_{23}+s_{12}s_{23}s_{31}e^{i\phi } & s_{23}c_{31} \\ 
s_{12}s_{23}-c_{12}c_{23}s_{31}e^{i\phi } & 
c_{12}s_{23}+s_{12}c_{23}s_{31}e^{i\phi } & -c_{23}c_{31}
\end{array}
\right)
\end{eqnarray}
(Our parametrization differs slightly from that of Chau and Keung in that we
take $\det V_{23}=\det V_{12}=-1$ in accordance with our earlier sign
choice.) Evidently, for the Ansatz in (\ref{theV}) and in (\ref{VGiunti}) $%
\theta _{31}=0$ and thus $V_{e3}=s_{31}e^{-i\phi }=0.$

If we take $V_{\text{exp}}^{\text{corr}}$ seriously, then a fit to $V_{{\rm %
angular}}$ gives 
\begin{equation}
\theta _{12}=0.617=35.4^{o};\theta _{23}=0.800=45.8^{o};\theta
_{31}=0.056=3.21^{o}.
\end{equation}
These values are extremely close by construction to the corresponding values
for the Ansatz $V_{HPSHZ}$ in (\ref{theV})

\begin{equation}
\theta _{12}=35.3^{o};\theta _{23}=45^{o};\theta _{31}=0^{o}.
\end{equation}

\smallskip

\smallskip

This discussion makes clear how easy it is to construct more theoretical
Ansatzes. Denote the three columns in the theoretical $V$ by $\vec{v}_{1},%
\vec{v}_{2},$ and $\vec{v}_{3}$ respectively. Pick one of the columns, call
it $\vec{v},$ compare with the corresponding column in $V_{\text{exp}}^{%
\text{corr}}$ and fit two of the components in $\vec{v}$ to some
combinations of (due to some possibly unjustifiable theoretical prejudice)
small integers and their square roots. The third component of $\vec{v}$ is
then fixed by orthonormality. Pick another column, call it $\vec{v}^{\prime
},$ and fit one of the components in $\vec{v}^{\prime }$ to some
combinations of small integers and their square roots. The other two
components of $\vec{v}^{\prime }$ are then fixed by orthonormality and by
orthogonality to $\vec{v};$ they emerge as solutions of a quadratic
equation. The last column is then fixed to be $\vec{v}^{\prime \prime }=\vec{%
v}\times $ $\vec{v}^{\prime }$ up to a sign.

\bigskip

{\bf {\large V. To Make Some Experimentalists Happy}}

\bigskip

A number of major experiments to detect $CP\;$\bigskip violation in neutrino
oscillation are being planned. Unfortunately, if $V_{e3}$ vanishes the
observable signal for $CP$ \bigskip violation also vanishes, as is shown
clearly in (\ref{cklike}). The vanishing of $V_{e3},$ however, is the one
common feature of $V_{HPSHZ}$ and $V_{G}.$ Thus, the experimentalists
involved have a human tendency to strongly dislike both $V_{HPSHZ}$ and $%
V_{G}.$

Suppose, in order to make some experimentalists happy, we try to construct
an alternative Ansatz $V_{H}$ with $(V_{H})_{e3}=1/5,$ close to its
experimental upper bound (\ref{exp}) of $0.22$. Of course at each step of
the construction of $V_{H}$ we will have to make an arbitrary choice. We
will choose what I consider to be the most reasonable. Suppose we want $%
(V_{H})_{\mu 3}$ and $(V_{H})_{\tau 3}$ to remain equal in magnitude. This
fixes the last column of $(V_{H})$ to be $\{1,$ $2\sqrt{3},-2\sqrt{3}\}/5.$
Note that $(V_{H})_{\mu 3}=2\sqrt{3}/5$ remains very close to $1/\sqrt{2},$
its value in both $V$ and $V_{G},$ since its square is equal to $12/25.$ Let
us next suppose $(V_{H})_{e2}=1/\sqrt{3}$ as suggested by experiment.
Orthogonality then fixes $(V_{H})_{\mu 2}$ and $(V_{H})_{\tau 2}$ to be $(%
\sqrt{141}-\sqrt{3})/12\sqrt{3}$ and $(\sqrt{141}+\sqrt{3})/12\sqrt{3}$
respectively. The first column is completely fixed to be the vector cross
product of the second and third columns. We obtain what some might consider
a rather ``ugly'' matrix 
\begin{equation}
V_{H}=\left( 
\begin{array}{rrr}
{-}\frac{1}{5}\sqrt{\frac{47}{3}} & {\frac{1}{\sqrt{3}}} & \frac{1}{5} \\ 
{\frac{5}{12}+}\frac{\sqrt{47}}{60} & {\frac{\sqrt{141}-\sqrt{3}}{12\sqrt{3}}%
} & \frac{2\sqrt{3}}{5} \\ 
{\frac{5}{12}-}\frac{\sqrt{47}}{60} & {\frac{\sqrt{141}+\sqrt{3}}{12\sqrt{3}}%
} & -\frac{2\sqrt{3}}{5}
\end{array}
\right)  \label{vh}
\end{equation}
Theoretically, even if we were optimistic enough to imagine that we could
derive $V$ or $V_{G}$ (from some discrete symmetry group\cite{lv} for
example) it is hard to imagine how we could possibly derive $V_{H}.$
However, $V_{H}$ provides a reasonable fit to experiment by construction;
numerically 
\begin{equation}
V_{H}=\left( 
\begin{array}{rrr}
{-0.792} & {0.577} & 0.200 \\ 
{0.531} & {0.488} & 0.693 \\ 
{0.302} & {0.655} & -0.693
\end{array}
\right)
\end{equation}
to be compared with $(\ref{correxpnum}).$ We note that 
\begin{equation}
d(V_{H})\equiv 3-Tr(V_{H}^{T}V_{\text{exp}}^{\text{corr}})\simeq 0.0214,
\end{equation}
about a factor of 2 larger than $d(V_{G}).$ This somewhat meaningless
exercise merely shows that it is perfectly consistent with present data to
have a large $e3$ matrix element, but it also illustrates the fact, which
the reader can convince himself or herself by playing around, that it is not
easy to come up with a ``nice'' theoretical Ansatz for the mixing matrix.
Orthogonality typically generates square roots of large integers, and by
going through a number of cases, I see that this usually happens unless $%
V_{e3}=0$ as in $V_{HPSHZ},$ as may be obvious to the reader.

\bigskip

{\bf {\large VI. The Value of }}$V_{e3}$

\smallskip

As mentioned earlier, the value of $V_{e3}$ is of crucial interest to some
experimentalists. In the last section we simply put in the value $1/5,$ and
in both the theoretical Ansatzes (\ref{theV}) and (\ref{VGiunti}) $V_{e3}$
is set to $0.$ We could perhaps turn things around and try to ``predict'' $%
V_{e3}.$

\smallskip We will try to find another Ansatz, which we will call $V_{C},$
by fitting 3 numbers from the first and second columns of $V_{\text{exp}}^{%
\text{corr}},$ and see what we get for $V_{e3}.$ For the reader's
convenience I reproduce $V_{\text{exp}}^{\text{corr}}$ here: 
\begin{equation}
V_{\text{exp}}^{\text{corr}}=\left( 
\begin{array}{lll}
-0.814 & 0.578 & 0.056 \\ 
0.436 & 0.545 & 0.716 \\ 
0.384 & 0.607 & -0.696
\end{array}
\right) .  \label{vexpagain}
\end{equation}
Let us look at column 2. As I already noted in constructing (\ref{theV}), $%
\frac{1}{\sqrt{3}}\simeq 0.577$ is very close to $(V_{\text{exp}}^{\text{corr%
}})_{e2}$ and so we set $(V_{C})_{e2}=\frac{1}{\sqrt{3}}.$ But in contrast
to the construction of (\ref{theV}), here we will take seriously the
apparent fact that $(V_{\text{exp}}^{\text{corr}})_{\mu 2}$ is somewhat
smaller than $(V_{\text{exp}}^{\text{corr}})_{e2}:$ with a bit of fiddling
we find that $\frac{2}{3}\sqrt{\frac{2}{3}}\simeq 0.544$ gives a good fit to 
$(V_{\text{exp}}^{\text{corr}})_{\mu 2}.$ Orthonormality then fixes $(V_{%
\text{exp}}^{\text{corr}})_{\tau 2}$ to be $\frac{1}{3}\sqrt{\frac{10}{3}}%
\simeq 0.609.$ Now that we have filled column 2 of $V_{C}$ we can only fix
one of the numbers in column 1: orthonormality of column 1 and orthogonality
between column 1 and column 2 will fix the other two. We observe that $\frac{%
2}{3}\sqrt{\frac{1}{3}}\simeq 0.385$ is very close to $(V_{\text{exp}}^{%
\text{corr}})_{\tau 1},$ so let's set $(V_{C})_{\tau 1}=\frac{2}{3}\sqrt{%
\frac{1}{3}}.$ We now have to solve a quadratic equation to determine $%
(V_{C})_{e1}$ and $(V_{C})_{\mu 1},$ and here comes a ``problem'': as in the
last section, there is no reason that the solution of a generic quadratic
equation would involve only small integers and the square root of small
integers. In fact, already $\sqrt{10}$ has appeared. Indeed, we obtain $%
(V_{C})_{e1}=-\frac{2}{51}(3\sqrt{26}+\sqrt{30})$ and $(V_{C})_{\mu 1}=\frac{%
1}{153}(27\sqrt{13}-8\sqrt{15}).$ The vector $\vec{v}_{3}$ in column 3 is
now determined to be the vector cross product of the vectors $\vec{v}_{1}$
and $\vec{v}_{2}$ in column 1 and column 2, and we obtain the Ansatz 
\begin{equation}
V_{C}=\left( 
\begin{array}{lll}
-\frac{2}{51}(3\sqrt{26}+\sqrt{30}) & \frac{1}{\sqrt{3}} & \frac{\sqrt{2}}{51%
}(-12+\sqrt{195}) \\ 
\frac{1}{153}(27\sqrt{13}-8\sqrt{15}) & \frac{2}{3}\sqrt{\frac{2}{3}} & 
\frac{2}{153}(27+2\sqrt{195}) \\ 
\frac{2}{3}\sqrt{\frac{1}{3}} & \frac{1}{3}\sqrt{\frac{10}{3}} & -\frac{1}{3}%
\sqrt{\frac{13}{3}}
\end{array}
\right) .  \label{VC}
\end{equation}
Thus, we ``predict'' $V_{e3}$ to be 
\begin{equation}
\frac{\sqrt{2}}{51}(\sqrt{195}-12)\simeq 0.0544677,  \label{ve3predict}
\end{equation}
which ``happens'' to be in a range encouraging to experimentalists.
Numerically,

\begin{equation}
V_{C}=\left( 
\begin{array}{lll}
-0.815 & 0.577 & 0.055 \\ 
0.434 & 0.544 & 0.718 \\ 
0.385 & 0.608 & -0.694
\end{array}
\right) .  \label{vcnum}
\end{equation}
By inspection we see that (\ref{vcnum}) gives a good fit to (\ref{vexpagain}
); this is confirmed by the group invariant distance between $V_{C}$ and $V_{%
\text{exp}}^{\text{corr}}$: 
\begin{equation}
d(V_{C})\equiv 3-Tr(V_{C}^{T}V_{\text{exp}}^{\text{corr}})\simeq 1.08\times
10^{-5}.
\end{equation}
Compare $d(V_{C})$ with $d(V)$ and $d(V_{G}).$

Of the three Ansatzes (I am discounting $V_{H})$ $V_{C}$ emerges as the
champion. However, most theorists, perhaps unjustifiably prejudiced against
large integers (they might pop out of string theory?), would probably regard 
$V_{C}$ as less elegant than the other two Ansatzes even though it gives a
better fit. Thus, in a sense, this exercise reinforces my belief in, or
rather, liking for the Ansatz $V$ in (\ref{theV}).

Whenever experimentalists refine the raw data in (\ref{exp}) the reader can
easily play this game. Extract some kind of mean value $V_{\text{mean}}$ and
then orthogonalize or unitarize it to an orthogonal or unitary $V_{\text{exp}%
}^{\text{corr}}.$ The proposed Ansatz should be judged by its group
theoretic distance to $V_{\text{exp}}^{\text{corr}}.$ This procedure is
further illustrated in the Appendix.

\smallskip

\bigskip

{\bf {\large VII. Neutrino Mass Matrix}}

\smallskip

\bigskip Neutrino oscillation experiments can only determine the absolute
value of the mass squared differences $\Delta m_{ij}^{2}\equiv $ $%
m_{i}^{2}-m_{j}^{2}$ rather than the three individual masses $m_{i}.$ At the
99.3\% confidence level $\Delta m_{ij}^{2}$ are determined to be $1.5\times
10^{-3}eV^{2}\leq |\Delta m_{32}^{2}|\leq 5.0\times 10^{-3}eV^{2}$ and $%
2.2\times 10^{-5}eV^{2}\leq |\Delta m_{21}^{2}|\leq 2.0\times 10^{-4}eV^{2}$%
, with the best fit values given by 
\begin{equation}
|\Delta m_{32}^{2}|=3\times 10^{-3}eV^{2}  \label{32}
\end{equation}
and 
\begin{equation}
|\Delta m_{21}^{2}|=7\times 10^{-5}eV^{2}  \label{21}
\end{equation}
We could have either the so-called normal hierarchy in which $%
|m_{3}|>|m_{2}|\sim |m_{1}|$ or the inverted hierarchy $|m_{3}|<|m_{2}|\sim
|m_{1}|.$ At present, we have no understanding of the neutrino masses just
as we have no understanding of the charged lepton and quark masses.

\bigskip Suppose we claim that we know the neutrino mixing matrix $V.$ As
before, call the three column vectors in the mixing matrix $\vec{v}_{i}$.
The neutrino mass matrix is then given by 
\begin{equation}
M=\sum_{i=1}^{3}m_{i}\vec{v}_{i}(\vec{v}_{i})^{T}.
\end{equation}
In other words, the 6 parameters of a real symmetric matrix have been
reduced to the 3 eigenvalues $\{m_{i}\}.$ In particular, if we believe in $%
V_{HPSHZ}$ we have 
\begin{equation}
M=\frac{m_{1}}{6}\left( 
\begin{array}{lll}
4 & -2 & -2 \\ 
-2 & 1 & 1 \\ 
-2 & 1 & 1
\end{array}
\right) +\frac{m_{2}}{3}\left( 
\begin{array}{lll}
1 & 1 & 1 \\ 
1 & 1 & 1 \\ 
1 & 1 & 1
\end{array}
\right) +\frac{m_{3}}{2}\left( 
\begin{array}{lll}
0 & 0 & 0 \\ 
0 & 1 & -1 \\ 
0 & -1 & 1
\end{array}
\right)  \label{massmatrix}
\end{equation}

In \cite{hz1}, in discussing the neutrino mass matrix, we proposed a basis
of 3 matrices other than those that appear in (\ref{massmatrix}). First, the
3 column vectors in $V$ are the eigenvectors of the matrix 
\begin{equation}
M_{0}=a\left( 
\begin{array}{lll}
2 & 0 & 0 \\ 
0 & -1 & 3 \\ 
0 & 3 & -1
\end{array}
\right)  \label{m0}
\end{equation}
with eigenvalues $m_{1}=m_{2}=2a,$ and $m_{3}=-4a.$ (The parameter $a$
merely sets the overall scale.) Thus, with $M_{0}$ as the mass matrix $%
\Delta m_{21}^{2}=0$ and this pattern reproduces the data $|\Delta
m_{21}^{2}|/|\Delta m_{32}^{2}|\ll 1$ to first approximation.

Because of the degeneracy in the eigenvalue spectrum, $V$ is not uniquely
determined. To determine $V,$ and at the same time to split the degeneracy
between $m_{1}\ $and $m_{2},$ we perturb $M_{0}$ to $M=M_{0}+\delta M_{D},$
with the `` democratic'' form 
\begin{equation}
\delta M_{D}=\varepsilon a\left( 
\begin{array}{lll}
1 & 1 & 1 \\ 
1 & 1 & 1 \\ 
1 & 1 & 1
\end{array}
\right) .
\end{equation}
The matrix $\delta M_{D}$ is evidently a projection matrix that projects the
first and third columns in $V$ to zero. Thus, the eigenvalues are given by $%
m_{1}=2a,m_{2}=2a(1+3\varepsilon /2),$ and $m_{3}=-4a,$ where to the lowest
order $\varepsilon =\Delta m_{21}^{2}/\Delta m_{32}^{2}$ and $a^{2}=\Delta
m_{32}^{2}/12$. Finally, to break the relation $|m_{3}|=2|m_{1}|\simeq
2|m_{2}|$ we can always add to $M$ a term proportional to the identity
matrix.

\bigskip

{\bf {\large VIII. Conclusion}}

\bigskip

In this note, we made the simple point that experimental data should be
summarized and parametrized by its deviation from some suitable theoretical
Ansatz, so that we are effectively dealing with an orthogonal or unitary
matrix close to the identity, which we can then study in the same spirit as
the Wolfenstein parametrization. Here for simplicity we have focussed on the
orthogonal case, but it is straightforward to generalize our discussion to
the unitary case by adding a phase. We also advocate using the group
invariant distance between two matrices to measure the goodness of fit.

\bigskip

{\bf {\large Acknowledgments\medskip }}

I would like to thank the various neutrino experimentalists I spoke with at
the Tropical Neutrino Conference held in Palm Cove, Australia, June 2003.
This work was supported in part by the National Science Foundation under
grant number PHY 99-07949.\medskip

\smallskip

{\bf {\large Appendix}}

\smallskip

\smallskip

In the Ansatz $V_{C}$ discussed in section VI, I first set column 2 in $%
V_{C} $ to 
\begin{equation}
\vec{v}_{2}\equiv \left( 
\begin{array}{l}
V_{e2} \\ 
V_{\mu 2} \\ 
V_{\tau 2}
\end{array}
\right) =\left( 
\begin{array}{r}
{\frac{1}{\sqrt{3}}} \\ 
\frac{2}{3}\sqrt{\frac{2}{3}} \\ 
\frac{1}{3}\sqrt{\frac{10}{3}}
\end{array}
\right) .  \label{v2}
\end{equation}
and $V_{\tau 1}$ to $\frac{2}{3}\sqrt{\frac{1}{3}}.$ I then determined $%
V_{e1}$ and $V_{\mu 1}$ in 
\begin{equation}
\vec{v}_{1}\equiv \left( 
\begin{array}{l}
V_{e1} \\ 
V_{\mu 1} \\ 
V_{\tau 1}
\end{array}
\right)
\end{equation}
by solving the quadratic equation resulting from the conditions $\vec{v}%
_{1}\cdot \vec{v}_{1}=1$ and $\vec{v}_{1}\cdot \vec{v}_{2}=0.$ Finally, 
\begin{equation}
\vec{v}_{3}\equiv \left( 
\begin{array}{l}
V_{e3} \\ 
V_{\mu 3} \\ 
V_{\tau 3}
\end{array}
\right)
\end{equation}
was determined by $\vec{v}_{3}=\vec{v}_{1}\times \vec{v}_{2}.$ ``Large''
integers and a non-zero value of $V_{e3}$ emerge.

Now suppose, instead of fixing $V_{\tau 1}$ in $\vec{v}_{1},$ I follow the
two alternate possibilities: (A) fix $V_{e1}$ to $-\sqrt{\frac{2}{3}}\simeq
-0.817,$ or (B) fix $V_{\mu 1}$ to $\frac{\sqrt{3}}{4}\simeq 0.433.$ Note
for comparison that $(V_{\text{exp}}^{\text{corr}})_{e1}=-0.814$ and $(V_{%
\text{exp}}^{\text{corr}})_{\mu 1}=0.436.$ For possibility (A) and
possibility (B), let me ask you three questions. (1) Does a non-zero value
of $V_{e3}$ emerge? (2) Do ``large'' integers appear? (3) How close is the
resulting Ansatz to $V_{\text{exp}}^{\text{corr}}?$

\smallskip The point is that, at least for me, it is not easy to answer
these questions without doing the computation (it is of course easy to write
a short program to perform the computation).

It turns out that 
\begin{equation}
V_{A}=\left( 
\begin{array}{lll}
-\sqrt{\frac{2}{3}} & \frac{1}{\sqrt{3}} & 0 \\ 
\frac{2}{3}\sqrt{\frac{1}{3}} & \frac{2}{3}\sqrt{\frac{2}{3}} & \frac{\sqrt{5%
}}{3} \\ 
\frac{1}{3}\sqrt{\frac{5}{3}} & \frac{1}{3}\sqrt{\frac{10}{3}} & -\frac{2}{3}
\end{array}
\right)
\end{equation}
has a ``nice'' form without ``large'' integers and predicts $V_{e3}=0,$ with
a distance from $V_{\text{exp}}^{\text{corr}}$ of 
\begin{equation}
d(V_{A})\equiv 3-Tr(V_{A}^{T}V_{\text{exp}}^{\text{corr}})\simeq 0.00484,
\end{equation}
better than $V_{G}$ and slightly worse than $V_{HPSHZ}.$ Meanwhile, $V_{B}$
contains huge integers and is too ``ugly'' to write down here, but predicts 
\begin{equation}
V_{e3}=\frac{1}{228}(27\sqrt{10}-2\sqrt{1338})\simeq 0.0536
\end{equation}
in a range encouraging to experimentalists$,$ and deviates from $V_{\text{exp%
}}^{\text{corr}}$ by only 
\begin{equation}
d(V_{B})\equiv 3-Tr(V_{B}^{T}V_{\text{exp}}^{\text{corr}})\simeq 1.78\times
10^{-5},
\end{equation}
much better than both $V_{G}$ and $V_{HPSHZ}$ and only somewhat worse than $%
V_{C}.$

It appears that the absence of large integers and $V_{e3}=0$ are correlated.
If we take $V_{\text{exp}}^{\text{corr}}$ seriously and if we are not
prejudiced against large integers we can get a better fit with a
non-vanishing $V_{e3}.$


\smallskip

\bigskip

\bigskip

\begin{references}
\bibitem{wolfpara}  L. Wolfenstein, Phys. Rev. Lett. 51, 1945 (1983).

\bibitem{bkaus}  P. Kaus and S. Meshkov, hep-ph/0211338.

\bibitem{xingpara}  Z. Z. Xing, hep-ph/0211465.

\bibitem{GonGar}  Concha Gonzalez-Garcia,
http://www.dpf2003.org/xx/neutrino/concha.pdf.

\bibitem{bgp}  J.N. Bachall, M.C. Gonzalez-Garcia and C. Pena-Garay,
hep-ph/0212147.

\bibitem{kay}  B. Kayser, hep-ph/0211134.

\bibitem{nieves}  J. F. Nieves and P. B. Pal, Phys. Rev. {\bf D36},
315(1987); S. Bilenky and S. Pascoli and S. Petcov, Phys. Rev. {\bf D64},
053010(2001).

\bibitem{sign}  A. de Gouvea, A. Friedland and H. Murayama, Phys. Lett. {\bf %
B 490}, 125(2000).

\bibitem{hz1}  X. G. He and A. Zee, Phys. Lett. {\bf B560}, 87(2003)
hep-ph/0301092.

\bibitem{hps}  P. F. Harrison, D.H. Perkins and W.G. Scott, Phys. Lett. {\bf %
B530}, 167(2002), hep-ph/0202074.

\bibitem{wolf78}  L. Wolfenstein, Phys. Rev. {\bf D18}, 958(1978).

\bibitem{21}  P.F. Harrison, D. H. Perkins and W.G. Scott, Phys. Lett. {\bf %
B458}, 79(1999); W.G. Scott, Nucl. Phys. Proc. Suppl. {\bf 85}, 177(2000),
hep-ph/9909431; P. F. Harrison and W.G. Scott, Phys. Lett. {\bf B535},
163(2002); P.F. Harrison and W.G. Scott, hep-ph/0302025.

\bibitem{22}  Z.-Z. Xing, Phys. Lett. {\bf B533}, 85(2002), hep-ph/0204049.

\bibitem{lv}  C. I. Low and R. R. Volkas, hep-ph/0305243.

\bibitem{giunti}  C. Giunti, hep-ph/0209103.

\bibitem{chaukeung}  L.-L. Chau and W.-Y. Keung, Phys. Rev. Lett. {\bf 53},
1802 (1984)
\end{references}
\end{document}